

      \parskip=0pt plus 1pt
      \def\chapter#1#2{\vskip1cm\line{\hfill{\bf#1\quad\uppercase{#2}}\hfill}
           \vskip1cm
           \headline={\ifodd\pageno{\hfil{\tentt#2\quad#1}}
                      \else{\tentt#1\quad#2\hfil}\fi}}
      
      \def\endchapter{\ifodd\pageno{\vfil\eject\chapter{}{}\line{}\vfil\eject}
                      \else{\vfil\eject}\fi}
      \def\section#1#2{\vskip1truecm\noindent{\bf#1\ #2}\vskip0.5truecm}\indent
      \def\subsection#1#2{\vskip1truecm\noindent{\it#1\
           #2}\vskip0.5truecm}\indent



      \def\d{\partial}

      \def\half{{\textstyle{1\over2}}}
      
      \def\IP#1#2{\langle\, #1\, |\, #2\, \rangle}
      
      \def\Lapop{\displaystyle{{\hbox to 0pt{$\sqcup$\hss}}\sqcap}}
      
      \def\lint{\int\nolimits}

      \def\ovr{\overline}
      \def\pb#1{\rlap{\lower1ex\hbox{$\leftarrow$}}#1{}}
      \def\pf#1{\rlap{\lower1ex\hbox{$\rightarrow$}}#1{}}
      
      \def\real{{\rm I\!R}}
      \def\to{\rightarrow}

      \def\ut#1{\rlap{\lower1ex\hbox{$\sim$}}#1{}}


 \def\CQG{Class. \&\ Quant.\ Grav.}
 \def\GRG{Gen.\ Rel. \&\ Grav.}
 \def\JMP{J.\ Math.\ Phys.}

 \def\PR{Phys.\ Rev.}
 \def\PRL{\PR\ Lett.}

 \hsize=6truein\overfullrule=0pt

\def\ca{{\cal A}} \def\cas{\ca^{(\star)}}
\def\vk{\vec{k}} \def\vx{\vec{x}} \def\vkp{\vec{k'}} 
\def\d{\partial}
%
\centerline{\bf Self Duality and Quantization}
\bigskip
\centerline{Abhay Ashtekar${}^{1,2}$, Carlo Rovelli${}^{3,4}$ and
Lee Smolin${}^{1}$}

\centerline{${}^1$ {\it Physics Department, Syracuse University, Syracuse, NY
13244- 1130, USA.}}
\centerline{${}^2$ {\it Theoretical Physics, Imperial College, London SW7 2BZ,
UK.}}
\centerline{${}^3${\it Physics Department, University of Pittsburgh,
Pittsburgh,
PA 15260, USA.}}
\centerline{${}^4${\it Dipartimento di Fisica, Universit\'a di Trento, 38050
Pavia, Italia.}}
\bigskip

Quantum theory of the free Maxwell field in Minkowski space is constructed
using a representation in which the {\it self dual} connection is diagonal.
Quantum states are now holomorphic functionals of self dual connections and a
decomposition of fields into positive and negative frequency parts is
unnecessary. The construction requires the introduction of new mathematical
techniques involving ``holomorphic distributions''. The method extends also to
linear gravitons in Minkowski space. The fact that one can recover the entire
Fock space --with particles of {\it both} helicities-- from self dual
connections alone provides independent support for a non-perturbative,
canonical quantization program for full general relativity based on self dual
variables.

\section{1}{Introduction}
Over the past three decades, Roger Penrose has provided us with several elegant
mathematical techniques to unravel the structure of zero rest mass fields [1].
In particular, we have learnt from him that the description of these fields is
especially simple if we decompose them using 2-component spinors. For spin-1
and spin-2 fields --which mediate all four basic interactions-- this amounts
to focussing on the eigenstates of the Hodge-duality operator. It is striking
indeed to see how, in the classical theory, the non-linear Yang-Mills and
Einstein equations simplify in the self dual (or anti-self dual) sector. The
richness of the mathematical structure of these solutions tempts one to
conjecture that the notion of self duality should play a significant role in
the quantization of these fields.

To see how this may come about concretely, let us first recall the relevant
features of the theory of fields satisfying free relativistic equations in
Minkowski space. The space of complex solutions to these equations
provides us with unitary representations of the Poincar\'e group [2].
These are classified by the values of the two Casimir operators, mass and spin,
or, if the mass is zero, helicity. Let us focus on Maxwell fields. Then the
construction yields four irreducible unitary representations consisting,
respectively, of positive frequency self dual fields, positive frequency
anti-self dual fields, negative frequency self dual fields and negative
frequency anti-self dual fields. All of them have zero mass. The first and the
fourth have helicity $+1$ while the second and the third have helicity $-1$
[3]. In the quantum theory, one generally chooses to work with the positive
frequency polarization: 1-photon states are represented by positive frequency
fields and more general quantum states, by entire holomorphic functionals on
the 1-photon Hilbert space. In this description, positive helicity photons
correspond to self dual fields and the negative helicity ones to anti-self dual
fields.

Note, however, that at least in principle one could have adopted another
strategy: one could have restricted attention just to self dual fields.
 \footnote{1}{Just as the description in terms of negative frequency fields
 simply mirrors the standard, positive frequency description,
 that in terms of anti-self dual fields would mirror the one in terms of
 self dual fields.}
{}From the result [3] on helicity of various sectors it would appear that,
at least apriori, such a description should be viable. The positive frequency
fields would now yield the helicity $+1$ photons and the negative frequency
ones, helicity $-1$ photons. The strategy seems attractive because
one would not have to {\it begin} by decomposing fields into positive and
negative frequency parts, an operation which has no counterpart beyond the
linear field theory in Minkowski space. Decomposition of fields into self dual
and anti-self dual parts, on the other hand, is meaningful both for the
non-linear gauge fields and general relativity. This then would be a concrete
way in which self duality could play a key role in quantization.

Why has this avenue not been pursued in the literature? As we shall see in some
detail, if one uses a self dual polarization in a straightforward way, one runs
into a problem: although the polarization is well-defined, unlike in the
positive frequency case, it fails to be K\"ahler. More precisely, the situation
is as follows. In the standard quantization procedure, one is naturally led to
define the inner product on positive frequency fields $F^+$ in terms of the
symplectic structure $\Omega$: $<F^+_1\>,\> F^+_2> := \Omega\>(\ovr{F^+_1}\>,
\>F^+_2)$. The same principles lead one, in the case of the self dual
polarization, to use the above expression for the inner product replacing only
the positive frequency fields by the self dual ones. However, this strategy now
fails: the resulting norm is no longer positive definite. The origin of the
problem is of course that in the self dual polarization, one works with both
positive and negative frequency fields and the inner product given above yields
negative norms on negative frequency fields. One may attempt to rescue the
situation by changing the inner product, introducing a minus sign {\it by hand}
on the negative frequency part of the self dual sector. This would of course
resolve the problem of negative norms. However, now a new problem arises: the
algebra of field operators is no longer faithfully represented on the resulting
Hilbert space! Thus, if one wishes to work in a representation in which the
self dual Maxwell connection is diagonal, one must modify the quantization
procedure. In particular, a new input is needed to select the appropriate inner
product.

The purpose of this paper is to supply the necessary modifications in the case
of the free Maxwell field.

For several decades now there has been available a fully satisfactory quantum
theory of photons in terms of positive frequency fields. It is therefore clear
that, were we interested {\it only} in Maxwell fields, there is really no need
for the alternate strategy mentioned above. The motivation for this work comes,
rather, from the quarters of quantum general relativity. Through his non-linear
graviton construction, Penrose [4] has taught us that unexpected
simplifications
arise when one deals with self dual gravitational fields in general relativity.
Einstein's field equations suddenly become more transparent, readily malleable
and fully manageable. While the chances of finding a general solution to the
full Einstein's equations still appear to be remote, for fifteen years we have
known that, inspite of all the non-linearities, the self dual sector is
completely integrable [4,5]! Therefore, it seems natural to base a
non-perturbative approach to quantum general relativity on this sector [6].
Such an approach is indeed being pursued vigorously and has already led to some
unexpected insights. (For recent reviews, see [7,8]). To have a greater
confidence in this approach, however, it is necessary to verify that the main
physical ideas and mathematical techniques it uses are viable in familiar
theories as well. It is in this spirit that we wish to recover here the
standard Fock description of photons working, however, in a representation in
which the self dual Maxwell connection is diagonal. A similar construction is
available also for linear gravitons [9]. However, in that case, a number of new
ideas come into play. The case of the Maxwell field has the advantage in that
we can focus just on one issue: Can one carry out quantization in a self dual
representation?

Section 2 is devoted to preliminaries. In the first part, 2.1, we outline the
quantization program we wish to follow. This is a simplified version of a more
general program that is being used for quantum general relativity [7]; we
have merely extracted the steps that are needed in the simpler, Maxwell case.
In
the second part, 2.2, we first recall the canonical framework underlying
Maxwell theory and then use the quantization program of section 2.1 to obtain
the precise statement of what is meant by a self dual representation.
Section 3 contains the main results. We find that the modes of the self
dual Maxwell field naturally split into two parts which can, at the end, be
identified with the two helicity states. (It is important to note that an
explicit decomposition of the field into its positive and negative frequency
parts is {\it not} carried out anywhere in the construction.) We show in
section 3.1 that quantization of positive helicity states is straightforward
within the framework of the program although the final description is
unconventional in certain respects. In section 3.2, we show that the negative
helicity states can also be quantized using the general framework of the
program. However, now, the steps involved are more subtle and require new
mathematical tools. The final picture is then summarized in section 3.3. We
conclude in section 4 by discussing some of the ramifications of these results.

In this paper we use a canonical approach based on space-like 3-surfaces.
An analogous treatment of the Maxwell field on null planes is given in
Josh Goldberg's contribution to this volume.

\goodbreak
\section{2}{Preliminaries}
Our primary aim in this paper is to use Maxwell fields as a probe to test
certain aspects of a non-perturbative approach [6-8] to quantum gravity based
on self dual fields. Therefore, we will closely follow the quantization program
developed there even though the steps involved may not appear to be the most
natural ones from the strict standpoint of the Maxwell theory. In the first
part of this section, we outline the general program emphasizing the points at
which new input is needed for quantization. This program is based on the
canonical quantization method and is applicable for a wide class of systems. In
the second part, we focus on the Maxwell field and construct the structures
needed in the program. We will then be able to give a detailed and precise
formulation of what is meant by a self dual representation. The problem of
constructing this representation will be taken up in the next section.

\goodbreak
\subsection{2.1}{The quantization program}
Consider a classical system with phase space $\Gamma$. To quantize the system,
we wish to follow an algebraic approach. We will proceed in the following
steps.
 \footnote{2}{For simplicity, we assume that there are no constraints.
 A more complete discussion of the program, including a treatment of
 constraints, is given in [7].}
\item{1.} Choose a subspace ${\cal S}$ of the space of complex valued
function(al)s on $\Gamma$ which is closed under the Poisson bracket operation
and which is large enough so that any well-behaved function(al) on $\Gamma$
can be expressed as (possibly a limit of) a sum of products of elements of
${\cal S}$. Elements of ${\cal S}$ are referred to as {\it elementary
classical variables} and are to have unambiguous quantum analogs. For a
non-relativistic particle moving in a potential, for example, $\Gamma$ is just
$\real^6$ and the elementary classical variables are generally taken to be
the three configuration variables $q^i$ and their conjugate momenta $p_i$.
\item{2.} Associate with each $f$ in ${\cal S}$, an abstract operator
$\hat{f}$. Even though at this stage there is no Hilbert space for them to
act upon, we will refer to the $\hat{f}$ as {\it elementary quantum operators}.
Consider the free associative algebra they generate and impose on it the
(generalized) canonical commutation relations (CCR):
    $$[\hat{f}, \hat{g}] = i\hbar\widehat{\{f,g\}},\qquad \forall f,g \in
        {\cal S}\>.\eqno(2.1)$$
Denote the resulting associative algebra by ${\cal A}$. For the
non-relativistic particle, (2.1) are just the standard Heisenberg commutation
relations and elements of ${\cal A}$ are simply sums of products of $\hat{q}^i$
and $\hat{p}_i$ with identifications implied by the CCR.
\item{3.} Introduce an involution operation, $\star$, on ${\cal A}$ by
first defining
   $$ (\hat{f})^\star = \hat{\bar{f}}, \qquad \forall f \in {\cal S},
       \eqno(2.2)$$
where $\bar{f}$ is the complex conjugate of the elementary classical variable
$f$, and extend the action of $\star$ to all of ${\cal A}$ by requiring
that it satisfy the three defining properties of an involution: i) $(\hat{A}
+\lambda \hat{B})^\star = \hat{A}^\star + \bar{\lambda}\hat{B}^\star$ ;
ii)$(\hat{A}\hat{B})^\star = \hat{B}^\star \hat{A}^\star$; and, iii)
$(\hat{A}^\star)^\star = \hat{A}$, where $\hat{A}$ and $\hat{B}$ are arbitrary
elements of ${\cal A}$ and $\lambda$ is any complex number. Denote the
resulting $\star$-algebra by ${\cal A}^{(\star)}$. Note that, at this stage,
$\cas$ is an abstract $\star$-algebra; the $\star$-operation does {\it not}
correspond to Hermitian conjugation on any Hilbert space. For the
non-relativistic particle, $\cas$ is obtained simply by making each
$\hat{q}^i$ and each $\hat{p}_i$ its own $\star$-adjoint.
\item{4.} Choose a linear representation of $\ca$ on a vector space $V$. The
$\star$-relations are ignored in this step. One simply wishes to incorporate
the linear relations between the operators and the CCR. For the
non-relativistic particle, one may choose for $V$ the space of smooth
functions $\Psi(\vec{q})$ with compact support on $\real^3$, represent $\hat
{q}^i$ by a multiplication operator and $\hat{p}_i$ by $i\hbar$ times a
derivative operator.
\item{5.} Introduce on $V$ an Hermitian scalar product $<\>,\>>$ by demanding
that the abstract $\star$-relations become concrete Hermitian-adjointness
relations:
$$ <\Psi\> ,\> \hat{A}\Phi>\> =\> <\hat{A}^\star \Psi\>, \> \Phi>\qquad
  \forall \hat{A}\in \ca,\> {\rm and}\> \forall \Psi, \Phi \in V. \eqno(2.3)$$
Note that (2.3) is now a condition on the choice of the inner product. The
Hiblert space ${\cal H}$ is obtained by taking the Cauchy completion of
$(V, <\>,\>>)$.

The program requires two external inputs: the choice of the space ${\cal S}$ in
step 1 and the choice of the representation in step 4. One may make ``wrong''
choices and find that the program cannot be completed (for examples, see [7]
and also section 3.2 below.) However, if the choices {\it are} viable-- i.e.,
if the program can be completed at all-- one is ensured the uniqueness of the
resulting quantum description in a certain well-defined sense [7]. For the
non-relativistic particle, for example, the choices we made are viable and step
5 does indeed provide the standard $L^2$-inner product on $V$. In the framework
of this program, the text-book treatment of free fields in Minkowski space can
be summarized as follows. The phase space $\Gamma$ can be taken to be the space
of (real) solutions to the field equations; smeared out fields provide
elementary variables; the $\star$-relations say that each smeared out field
operator is its own star; the representation space $V$ is the space of
holomorphic functionals of positive frequency classical fields; field operators
are represented as sums of multiplication (creation) and derivative
(annihilation) operators and the unique inner product which realizes the
$\star$-relations is given by the Poincar\'e invariant Gaussian measure on the
space of positive frequency fields. (See, e.g., [10]).

\goodbreak
\subsection{2.2}{Self dual variables for the Maxwell field}
Let us begin with a brief summary of the standard phase space formulation of
Maxwell fields. Denote by $\Sigma$ a space-like 3-plane in
Minkowski space. Thus, $\Sigma$ is topologically $\real^3$ and is equipped with
a flat, positive definite metric $q_{ab}$.  The configuration variable for
the Maxwell field is generally taken to be the connection 1-form $A_a(\vx )$
--the vector potential for the magnetic field-- on $\Sigma$. Its canonically
conjugate momentum is the electric field $E^a(\vx )$ on $\Sigma$. The
fundamental Poisson bracket is:
$$\{A_a(\vx ),\>E^b(\vec{y})\} = \delta_a^b \delta^3(\vx, \vec{y}).\eqno(2.4)$$
The system has one first class constraint, $\d_aE^a(\vx) =0$. One can therefore
pass to the reduced phase space by fixing the gauge. For simplicity, let us
choose this avenue. The true dynamical degrees of freedom are then
contained in the pair $(A^T_a(\vx ), E^a_T(\vx ))$ of transverse (i.e.,
divergence-free) vector fields on $\Sigma$. Denote by $\Gamma$ the
phase space spanned by these fields; now there are no constraints and we are
working only with the true degrees of freedom. On $\Gamma$, the only
non-vanishing fundamental Poisson bracket is:
   $$\{A^T_a(\vx ),\>E^b_T(\vec{y})\} = \delta_a^b \delta^3(\vx ,\vec{y})
  - \triangle^{-1} \d_a\d^b \delta^3(\vec{x}, \vec{y})\>,\eqno(2.5)$$
where $\triangle$ is the Laplacian operator compatible with the flat metric
$q_{ab}$. It is convenient --although by no means essential-- to work in the
momentum space. Then, the true degrees of freedom are contained in the new
dynamical variables
$q_j(\vk),\>p_j(\vk)$ with $j = 1,2$ :
 $$\eqalign{ A_a^T(\vx)&= {1\over (2\pi)^{3/ 2}}\lint d^3\vk\,
  e^{i\vk\cdot\vx}\, (q_1(\vk)m_a(\vk) + q_2(\vk)\ovr{m}_a(\vk))\cr
  E^a_T(\vx)&= {1\over (2\pi)^{3/ 2}}\lint d^3\vk\,
  e^{i\vk\cdot\vx}\, (p_1(\vk)m^a(\vk) + p_2(\vk)\ovr{m}^a(\vk)),\cr}
\eqno(2.6)$$
where $\sqrt{2}m_a =\d_a\theta + i \sin\theta\> \d_a\phi$ is a complex vector
field in the momentum space which is transverse, $m_a(\vk )k^a = 0$, and
normalized so that $m_a(\vk )\ovr{m}^a(\vk ) = 1$. The Poisson brackets (2.5)
are equivalent to:
 $$\{q_i(-\vk ), \>p_j(\vkp )\} = -\delta_{ij}\>\delta^3(\vk ,\vkp ),
\eqno(2.7)$$
while the fact that $A_a^T(\vx )$ and $E^a_T(\vx )$ are real translates to
the conditions:
   $$ \ovr{q}_j(\vk ) = q_j(-\vk ) \quad {\rm and} \quad
      \ovr{p}_j(\vk ) = p_j(-\vk )\>.\eqno(2.8)$$

We are interested in using a self dual representation. Let us therefore
first construct the self dual connection from the pair $(A^T_a(\vx ), \>
E^a_T(\vx ))$. If we denote by $d_a^T(\vx )$ the transverse vector potential of
the {\it electric field},
   $$d^T_a(\vx ) = {1\over (2\pi)^{3/ 2}}\lint {d^3\vk \over{|\vk|}}\>
      e^{i\vk\cdot\vx}\, \left( p_1(\vk)\> m_a(\vk) -
       p_2(\vk)\>\ovr{m}_a(\vk)\right)\>,\eqno(2.9)$$
the self dual connection is given simply by:
       $$ {}^+\!A^T_a(\vx ) = -A^T_a(\vx ) + i d^T_a(\vx ).\eqno(2.10)$$
Following the procedure used in general relativity [6], we now want to use the
pair $({}^+\!A_a^T(\vx ), E^a_T(\vx ))$ as our basic variables. From the
viewpoint of the Maxwell theory, this choice is rather strange. However, it is
in terms of the analogous ``hybrid'' canonical  variables --one of which is
complex and the other real-- that the non-perturbative quantization program for
full, non-linear relativity is most easily formulated. Therefore, here we
will work with this unconventional choice. Following (2.9), let us expand
${}^+\!A_a^T(\vx )$ in terms of its Fourier components. We have:
    $${}^+\!A^T_a(\vx ) = {1\over (2\pi)^{3/ 2}}\lint {d^3\vk\over{|\vk|}}\>
      e^{i\vk\cdot\vx}\, \left(z_1(\vk)\> m_a(\vk) -
      z_2(\vk)\>\ovr{m}_a(\vk)\right)\>,\eqno(2.11)$$
with
$$ z_1(\vk ) = -|\vk |q_1(\vk ) +i p_1(\vk ) \quad {\rm and} \quad
   z_2(\vk ) = |\vk |q_2(\vk ) + ip_2(\vk ).\eqno(2.12)$$
In terms of these dynamical variables, the basic Poisson brackets are given by:
  $$ \{q_i(-\vk ), \>z_j(\vkp ) \} = -i \delta_{ij}\delta^3(\vk, \vkp ),
   \eqno(2.13)$$
and the ``reality conditions'' (2.8) become:
$$\eqalignno{\ovr{q}_j(\vk ) &= q_j(\vk )\qquad {\rm and}\cr
  \ovr{z}_1(\vk ) = - z_1 (-\vk) - 2|\vk | q_1(-\vk )\>, &\qquad
  \ovr{z}_2(\vk ) = - z_2 (-\vk) + 2|\vk | q_2(-\vk )&(2.14)\cr}$$

To summarize, our basic dynamical variables will be $(z_j(\vk ),
\>q_j(\vk ))$. They satisfy the Poisson bracket relations (2.13) and the
reality conditions (2.14). The Hamiltonian for the Maxwell theory,
$H := \lint_\Sigma d^3x\> (E^T\cdot E^T + B^T\cdot B^T)$ (where $B^T$ is the
magnetic field), can now be expressed as:
$$H = \lint d^3\vk\> \sum_j\> \ovr{z}_j(\vk ) z_j(\vk )\>, \eqno(2.15)$$
where $\ovr{z}_j(\vk )$ can be be regarded as functionals of $z_j(\vk )$ and
$q_j(\vk )$ given by  (2.15).

With this machinery at hand, we can now give a precise formulation of
the problem we want to analyze. In the quantization program, we wish to
use $z_j(\vk )$ and $q_j(\vk )$ as the {\it elementary classical
variables}. We can then carry out steps 2 and 3 of the program in a
straightforward fashion and arrive at a $\star$-algebra $\cas$. We
again need new input in step 4. We want to use a representation in which the
self dual connection --and hence the operators $\hat{z}_j(\vk )$-- are
diagonal. The obvious choice is to use for the vector space $V$ the space of
polynomials $\Psi(z_j(\vk ))$ (which are, in particular,
entire holomorphic functionals) and represent the $\hat{z_j}$ by multiplication
operators. The representation of $\hat{q}_j(\vk )$ is then dictated by the
generalized CCR that result from (2.13). This is the self dual representation
we are seeking. The key questions now are: Can step 5 be carried out to
completion? Does there exist an inner product which implements conditions
(2.3) which arise from the reality conditions (2.14)? Is the inner product
unique? Are the resulting Hilbert spaces large enough to accommodate the two
helicities of photons? And, finally, is the resulting quantum description
equivalent to the standard Fock theory?
\vfill\eject

\goodbreak
\section{3}{Quantum Theory}
The form of the Hamiltonian (2.14) suggests that the Maxwell field can be
regarded as an assembly of harmonic oscillators, containing two oscillators
(labelled by $j$) per momentum vector $\vk$. Note furthermore, that the
oscillators with $j=1$ are completely decoupled from those with $j=2$; each
set is separately closed under the Poisson bracket relations (2.13) and the
reality conditions (2.14). Therefore, we can simplify our task by first
examining each set separately and {\it then} combine the results we obtain.

Let us begin by reviewing [7] the situation with a single harmonic oscillator.
The phase space $\Gamma$ is now 2-dimensional, co-ordinatized by the real
functions $q$ and $p$. Following equation (2.7), let us choose the fundamental
Poisson bracket to be $\{q, p\} = -1$. The analog of the complex, self dual
variable is $z = q+ip$ (see Eq. 2.10). The idea now would be to regard $(z,q)$
as the elementary classical variables. Together with constants, they are indeed
closed under the Poisson bracket
               $$ \{q, \>z\} = -i\>, \eqno(3.1)$$
as well as the reality conditions
    $$ \ovr{q} = q \qquad {\rm and} \qquad \ovr{z} = -z +2q\>.\eqno(3.2)$$
The Hamiltonian $H := q^2 +p^2 $ now becomes:
    $$H = \ovr{z}z \equiv (-z +2q) z\>.\eqno(3.3)$$
Let us compare this structure with the one we encountered in section 2.2. If we
let $q$ and $z$ here be, respectively, the analogs of $q_j(-\vk )$ and $z_j(\vk
)$ of section 2.2, we find that for $j=2$ the two sets are {\it completely}
analogous. For $j=1$, however, there is a discrepancy: while the Poisson
brackets and the Hamiltonians match, the reality conditions differ by a sign in
one of the terms. This difference will turn out to play a crucial role in the
implementation of the quantization program.

We begin in sub-section 3.1 with the simpler, $j=2$ case. We carry out the
quantization program of section 2.1 step by step for the harmonic oscillator
described by equations (3.1)-(3.3). In the second part, 3.2, we examine
the ramifications of the sign discrepancy in the reality condition for the
$j=1$ modes. We will find that this sign difference forces one to enlarge
the framework and allow as states {\it holomorphic distributions}.
In the last sub-section, 3.3, we collect all these results and present a
coherent quantum description of the Maxwell field in the self dual
representation.

\goodbreak
\subsection{3.1}{The $j=2$ modes} The first step in the quantization program of
section 2.1 is the introduction of the space ${\cal S}$ of elementary classical
variables. For the harmonic oscillator considered above, the natural choice is
the complex vector space spanned by the functions $1, z, q$ on the real,
2-dimensional phase space $\Gamma$: This space is closed under the Poisson
bracket operations and clearly ``large enough'' since it provides a (complex)
co-ordinatization of $\Gamma$. To generate the algebra $\ca$, introduce, first,
the elementary quantum operators, $\hat 1, \hat{z},\hat{q}$, and on the
collection of their formal sums of formal products, impose the CCR:
$$[\hat{q},\>\hat{z} ] = i\hbar\widehat{\{q,z\}} \equiv \hbar\>.\eqno(3.4)$$
Using Eq. (3.2), the $\star$-relations are also straightforward to impose. Set:
  $$\hat{q}^\star = \hat{q} \qquad {\rm and} \qquad \hat{z}^\star =
    -\hat{z} + 2 \hat{q}\>.\eqno(3.5)$$
It is easy to check that the resulting $\star$-algebra $\cas$ is
isomorphic to the standard $\star$-algebra constructed from the operators
$ \hat{1}, \hat{q}, \hat{p}$ that one finds in text-books.

The next step is to find a representation of this algebra. It is here that the
presence of the hybrid variables $(z,q)$ suggests a new avenue. Let $V$ now be
the vector space of entire holomorphic functions $\Psi(z)$ and let the concrete
operators representing $\hat{q}$ and $\hat{z}$ be
  $$\hat{q}\cdot \Psi(z) = \hbar\>{d\Psi (z)\over dz} \quad\hbox{ and }\>
    \quad \hat{z}\cdot \Psi (z) = z\Psi (z)\>,\eqno (3.6)$$
so that the canonical commutation relations(3.4) are satisfied. Now, the
question is whether the last step, 5, of the quantization program can be
carried out successfully: Is there is an inner product on $V$ which realizes
the $\star$-relations (3.5)? Let us begin by introducing a general measure
$\mu(z,\bar{z})$ on the complex $z$-plane on which the wave functions are
defined and set the inner-product to be:
  $$\IP{\Psi (z)}{\Phi (z)} \>= i\lint\>dz\wedge d\bar{z}\>
     \mu (z, \bar{z})\>\ovr{\Psi (z)}\> \Phi (z)\>.\eqno(3.7)$$
(The factor of $i$ arises because $dp\wedge dq \equiv\Omega = 2i dz\wedge
d\ovr{z}$.) Positivity of norms requires that $\mu (z, \bar{z})$ must be real.
This condition ensures that the requirement that $\hat q$ is its own Hermitian
adjoint is satisfied if one chooses$\mu$ of the form $\mu\equiv
\mu(z+\bar{z})$. It only remains to impose the $\star$-relation on $\hat{z}$ as
a condition on the choice of the inner product. It turns out that this
condition now determines the form of $\mu$ {\it completely}! Upto an overall
multiplicative constant, $\mu$ is given by: $\mu (z,\bar{z}) =
\exp(-\textstyle{1\over 4\hbar}(z+\bar{z})^2)$. Thus, the Hilbert space of
quantum states consists of entire holomorphic functions of $z$ which are
normalizable with respect to the inner product:
$$\IP{\Psi (z)}{\Phi (z)}\>= i\lint\>dz\wedge d\bar{z}\>
   e^{-{1\over 4\hbar}(z+\bar{z})^2}\> \ovr{\Psi (z)}\> \Phi (z)\>.\eqno(3.8)$$
Note that there is freedom to add to the expression of the operator $\hat{q}$
any holomorphic function of $z$; this addition will not alter the commutation
relations. It is easy to work out the change in the measure caused by this
addition and show that the resulting quantum theory is unitarily equivalent
to the one obtained above.

The question now is whether the space of normalizable states is ``sufficiently
large". The simplest way to analyze this issue is to relate our Hilbert
space to the one used in the Bargmann quantization [11] of the harmonic
oscillator. Given a $\Psi(z)$ in our Hilbert space, set $f(z)= \exp
(-\textstyle{z^2\over 4\hbar})\>\Psi(z)$. Then, the finiteness of the norm
of $\Psi (z)$ is equivalent to:
   $$i\!\lint\> dz\wedge d\bar{z}\> e^{-{\textstyle{z\ovr{z}\over 2\hbar }}}\>
        |f(z)|^2 < \infty\>. \eqno(3.9)$$
Note that the left hand side is precisely the norm of $f(z)$ in the Bargmann
Hil
bert
space! (In particular, the integral converges for all polynomials $f(z)$.)
Thus, there is a 1-1 correspondence between our quantum states $\Psi(z)$ and
the Bargmann states $f(z)$; the space of normalizable states is indeed
``large enough''. Let us translate the action of the operators defined
in (3.6) to the space of Bargmann states $f(z)$, using the unitary transform
$\Psi(z)\mapsto f(z)=\exp(-\textstyle{z^2\over 4\hbar})\Psi(z)$. We find:
  $$\hat{q}\cdot f(z) = \hbar\>{df (z)\over dz} + {z\over 2}
    f(z)\quad \hbox{ and }\quad\hat{z}\cdot f(z) = z f(z)\>.\eqno(3.10)$$
Equations (3.10) are precisely the expressions of the operators $\hat q$ and
$\hat z$ in the Bargmann representation. Thus, the representation we
constructed using the hybrid $(q,z)$-variables in the quantization
program is unitarily equivalent to the Bargmann representation. Finally,
note that in both these representations $\hat{z}$ is the creation operator
and $\hat{z}^\star$ is the annihilation operator.

We conclude this sub-section with a remark relating our $z$-representation
to the standard Schr\"odinger representation of the quantum oscillator.
Our choice of $V$ and the representation (Eq. 3.6) was motivated by the
fact that $z$ is complex and $\hat{q}$ and $\hat{z}$ satisfy the canonical
commutation relations. Note, however, that one can also arrive at this choice
systematically [12] using the fact that the passage from $(q,p)$ to $(q,z)$
corresponds to a simple canonical transformation $(q, p) \to (q, dF/dq + ip)$
with generating function $F(q) = \textstyle{1\over 2}q^2$. Had we used the pair
$(q,p)$ as our basic variables, the algebraic quantization program would have
led us, as indicated in section 2.1, to the Schr\"odinger representation of the
harmonic oscillator. In this picture, the states are represented by
square-integrable functions $\psi (q)$ of the real variable $q$ and the basic
operators are given by $\hat{q}\cdot\psi (q) = q\psi (q)$ and
$\hat{p}\cdot\psi(q) = i\hbar\>(d\psi/dq)$. The canonical transformation now
lets us pass to the new ``momentum'' or $z$-representation via the usual
transform between the configuration and the momentum representations:
 $$\Psi(z) := \lint dq\>\> \exp\left({zq\over \hbar}-{{q^2}\over 2\hbar}
     \right)\> \psi (q)\>.\eqno(3.11)$$
The function $\Psi(z)$ is clearly holomorphic and, given a square-integrable
$\psi (q)$, the integral converges for all (complex values of) $z$. Thus, the
result of the transform of any Schr\"odinger wave function is an entire
holomorphic function in the $z$-representation. Using the expressions of the
operators $\hat{q}$ and $\hat{p}$ in the Schr\"odinger representation, and the
definition $\hat{z} = \hat{q} +i \hat{p}$ of  $\hat{z}$, we can now transform
the operators $\hat{q}$ and $\hat{z}$ from the $q$ to the $z$-representation.
The result is precisely Eq. (3.6).

\goodbreak
\subsection{3.2}{The $j=1$ modes}
In terms of $(z_j(\vk ), q_j(\vk ))$, the only difference in the $j=1$ and
$j=2$ modes is in the reality conditions. Let us therefore consider again
a single simple harmonic oscillator with hybrid phase space variables
$(z,q)$, proceed as in section 3.1 to construct the quantum algebra $\ca$
using the CCR (3.4), but introduce the $\star$-relation via:
  $$ \hat{q}^\star = \hat{q} \qquad {\rm and} \qquad
      \hat{z}^\star = -\hat{z} - 2\hat{q}\>.\eqno(3.12)$$
The only difference between (3.12) the $\star$-relations (3.5) of section 3.1
is in the sign of the very last term. Using these new $\star$-relations, we can
complete step 3 of the program and obtain a $\star$-algebra $\ovr{\cal
A}^{(\star)}$. Note that $\cas$ and $\ovr{\cal A}^{(\star)}$ are constructed
from the same associative algebra ${\cal A}$; the difference is only in the
involution operation $\star$. Since the $\star$-relations are ignored in the
step 4 of the program, we can attempt to use the same strategy as in
section 3.1. Let us then choose $V$ to consist of entire holomorphic functions
of $z$ and represent the operators via (3.6). Then, the canonical commutation
relations (3.4) are satisfied and we have a representation of the quantum
algebra $\ca$. Our final task is to introduce on $V$ an inner product so that
the $\star$-relations (3.12) become concrete Hermitian-adjointness relations on
the resulting Hilbert space. As before, let us first make the ansatz (3.7) and
then attempt to determine the measure $\mu(z,\ovr{z})$ using (3.12). As before,
the measure is uniquely determined. However, since there is a change in sign in
the reality condition, the sign in the exponent of the measure is now the
opposite of what it was in section 3.1. We obtain:
     $$\mu(z,\ovr{z}) = \exp \left(+{\textstyle{1\over 4\hbar}}
            (z+\ovr{z})^2\right). \eqno(3.13)$$
Consequently, the arguments that led us in section 3.1 to the conclusion that
the Hilbert space of normalizable states is infinite dimensional (and naturally
isomorphic to the Bargmann Hilbert space), now implies that there are {\it no}
(non-zero) entire holomorphic functions which are normalizable with respect to
the inner product of (3.13)! Thus, the change in sign in the reality conditions
make a crucial difference: a new strategy is now needed in the choice of
the linear representation.
  \footnote{3}{The new strategy is motivated by the transform from the $q$ to
   the $z$-representation discussed at the end of section 3.1. Note also that
   the choice of the
   vector space $V$ we are about to introduce would be necessary also in the
   Bargmann quantization, had the symplectic structure been of opposite sign,
   or, alternatively, if the symplectic structure had been the same but we had
   used anti-holomorphic wave functions. In either case, we would have found
  that the measure needed to ensure the correct reality conditions is
  $\exp(+ {\textstyle{z\ovr{z}\over 2\hbar}})$, whence no entire holomorphic
  function would have been normalizable. We would then have to use for states
  the holomorphic distributions introduced below.}%

A solution to this problem is suggested by the following considerations.
A simple calculation shows that the change in the sign in the reality
conditions amounts to exchanging the creation and annihilation
operators. Thus, while the operator $\hat{z}$ served as the creator in section
3.1, if we can complete the quantization program, it would now serve as the
annihilation operator. It must therefore map the vacuum to zero. This suggests
that, if the program is to succeed, we need to represent the vacuum by a
``holomorphic, delta distribution'' $\delta(z)$. Excited states can then
be built by acting on this vacuum repeatedly by $\hat{q}$.

Since the meaning of these holomorphic distributions is not apriori clear, let
us make a brief detour to introduce some mathematical techniques that are
needed. Let us begin by defining the holomorphic {\it generalized function}
--or
{\it distribution})-- $\delta(z)$. It will be the complex linear mapping from
the space of functions of the type $\sum f_i(z)g_i(\ovr{z})$, where $f_i(z)$
are entire holomorphic functions and $g_{i}(\ovr{z})$ are entire
anti-holomorphic functions, to the space of entire anti-holomorphic functions,
given by:
  $$\delta(z)\circ \sum_i f_i(z)g_i(\ovr{z}) = \sum_i f_i(0)g_i(\ovr{z}).
  \eqno(3.14)$$
Next, we can define the anti-holomorphic distribution $\delta(\ovr{z})$
simply by taking the complex conjugate of $\delta(z)$. This new distribution
has the action:
  $$\delta(\ovr{z})\circ \sum_i f_i(z)g_i(\ovr{z}) = \sum_i f_i(z)g_i(0).
          \eqno(3.15)$$
{}From these two basic distributions, we can construct others. The product of a
polynomial $a(z,\ovr z)$ with a distribution ${\cal F}(z, \ovr{z})$ will be a
new distribution, given by:
  $$[a(z,\ovr z){\cal F}(z, \ovr{z})]\circ\sum_if_i(z)g_i(\ovr z):=
   {\cal F}(z,\ovr{z})\circ
  \>a(z,\ovr{z})\sum_if_i(z)g_i(\ovr z)\>.\eqno(3.16)$$
Finally, using the Leibnitz rule as a motivation, we define the derivative of a
distribution ${\cal F}(z)$, as
 $$\bigg[{d\over dz}{\cal F}(z)\bigg]\circ\sum_i f_i(z)g_i(\ovr z):={d\over
dz}\left({\cal F}(z)\circ\sum_if_i(z)g_i(\ovr z)\right) - {\cal
F}(z)\circ{d\over dz}\>\sum_if_i(z)g_i(\ovr z).\eqno(3.17)$$
The derivative with respect to $\ovr{z}$ is defined similarly. As an example,
let us compute the derivatives of $\delta(z)$. We have:
  $${d\over d\ovr{z}}\delta(z) = 0, \quad{\rm and} \quad
    \left[{d\over dz}\delta(z)\right]
  \circ \sum_i f_i(z)g_i(\ovr{z}) = -\sum_i {df_i(z)\over dz}\bigg|_{z=0}g_i(
  \ovr{z}). \eqno(3.18)$$
Thus, $\delta (z)$ {\it is ``holomorphic'' and its derivative with
respect to $z$ is a distribution with the expected property.} Finally, we
notice that the {\it product} of the two distributions (3.14) and $(3.15)$ is
well-defined; it is just the two dimensional $\delta$-distribution and
therefore admits the standard integral representation:
$$\eqalign{[\delta(z)\delta(\ovr{z})]\circ \sum_i f_i(z)g_i(\ovr{z}) &=
\sum_i f_i(0)g_i(0)\cr
&\equiv\lint dq\wedge dp\> \delta^2(q,p;0,0)\>\sum_i f_i(z)g_i(\ovr{z}),\cr}
\eqno(3.19)$$
where, we have used $z=-q+ip$. Thus, one can regard $\delta(z)$ as the
``holomorphic square-root'' of the standard 2-dimensional $\delta$-distribution
on the 2-plane, picked out by the complex structure.

With this machinery at hand, let us proceed with the quantization program.
In step 4 of the program let us use, as the carrier space for the
representation, the space $V$ spanned by the holomorphic distributions of the
type $\Psi(z) =\sum (a_n(z))\>( d^n\delta(z)/dz^n)$ where each $a_n(z)$ is a
polynomial in $z$. Then, we can continue to represent the operators $\hat{q}$
and $\hat{z}$ by (3.6).  Let us define an inner product on $V$ via:
$$\eqalign{\IP{\Psi (z)}{\Phi, (z)}:=&\ovr{\Psi (z)}\>\Phi (z)\circ\mu \cr
   =& i\lint dz\wedge d\ovr{z}\> \mu(z,\ovr{z})\>\ovr{\Psi (z)}\>\>
   {\Phi (z)}\cr}\eqno(3.20)$$
where $\mu=\mu(z,\ovr z)$ is a measure to be determined by the reality
conditions and where, in the second step, we have used the integral
representation (3.19) of the product of holomorphic and anti-holomorphic
distributions. Thus, the ansatz is formally the same as the one used earlier.
Furthermore, the previous calculations go through step by step because they
only use the fact that the states are holomorphic, i.e., are annihilated by the
operator $d/d\ovr{z}$, and the measure $\mu(z,\ovr{z})$ is therefore again
given by (3.13). However, now, the integral does converge because of the
presence of $\delta$-distributions in the expressions of our states. Thus, the
inner product is well defined for all elements of $V$. The full  Hilbert space
${\cal H}$ is obtained just by Cauchy completion. Note incidently that while
we began with the delta-distributions with support at $z=0$, the Cauchy
completions includes states with support at other point. $\Psi(z)=
\delta(z,z_0)$, for example, belongs to ${\cal H}$ and represents a coherent
state.

As  is expected from our motivating remarks, in this representation, it is the
{\it annihilation} operator that is represented by $\hat{z}$. The vacuum state
is simply the normalized state $\Psi_0(z) = \delta(z)$. An orthogonal basis in
the Hilbert space is provided by the states $d^n\delta(z)/dz^n$. (Thus, we
could also have let the representation space $V$ to be the linear span of
states of the type $\sum a_n \>(d^n\delta(z)/dz^n)$, where $a_n$ are
constants.) The Hamiltonian is given by
$$\hat{H} = \half (\hat{z}^\star\hat{z} + 1)
  \equiv - \half \left( ({z}+2\hbar{d\over dz}){z}-1\right).
  \eqno(3.21)$$
Finally, note that, inspite of the appearance of distributions, this
representation does diagonalize the operator $\hat{z}$; it acts as the
multiplication operator.

To conclude this sub-section, we wish to point out that there is in fact an
alternate strategy available to quantize the harmonic oscillator with the
present reality conditions. Recall that the quantization program requires two
new inputs: the choice of the space ${\cal S}$ of elementary classical
variables in the first step and of the representation of the algebra $\ca$ in
the fourth step. In this sub-section, we used the same elementary variables as
in 3.1 and changed the carrier space $V$ of the representation of $\ca$ to
accommodate the new reality conditions. Alternatively, we could have changed
the space ${\cal S}$ itself in step 1. Let us choose $(q, \ovr{z})$ as the
elementary variables in place of $(q, z)$. Then, one can in fact proceed
exactly as in section 3.1, replacing $z$ everywhere by $\ovr{z}$. The program
can be completed without recourse to distributions --the states are just
polynomials in $\bar{z}$-- and yet the resulting description is equivalent to
the one we have obtained here. However, in this representation, it is
$\hat{\bar{z}} \equiv \hat{z}^\star$ that is diagonal rather than $\hat{z}$!
Thus, although this strategy is viable, it would have led us to the {\it
anti}-self dual representation of $j=1$ modes in the Maxwell theory. We are
led to consider distributions precisely because we want to retain the self dual
representation for the $j=1$ modes as well.

\goodbreak
\subsection{3.3}{Self dual representation}
Let us now combine the results of the last two sub-sections to construct
the self dual representation for the quantum Maxwell field. This is the
representation in which the self dual connection ${}^+\!\hat{A}^T_a(\vx )$
--or, equivalently, $\hat{z}_j(\vk )$-- is diagonal.

As discussed in the beginning of this section, we are led to use $z_j(\vk ),
q_j(\vk )$ as the elementary classical variables in the quantization program
of section 2.1. The elementary quantum operators are then $\hat{z}_j(\vk ),
\hat{q}_j(\vk )$ and the algebra $\ca$ is generated by their formal sums of
formal products subject to the CCR
    $$ [\hat{q}_i(-\vk ) , \hat{z}_j(\vkp )] =
        \delta_{ij}\>\hbar \delta^3(\vk ,\vkp )\>\eqno(3.22)$$
which mirror the Poisson bracket relations (2.13). The next step is the
introduction of the abstract $\star$-relations. The reality conditions (2.14)
lead us to the relations:
    $$ \eqalign{\hat{q}_j^\star &= \hat{q}_j\qquad {\rm and}\cr
(\hat{z}_1(\vk ))^\star = - \hat{z}_1 (-\vk) -& 2|\vk | \hat{q}_1(-\vk )
\quad (\hat{z}_2(\vk )^\star = - \hat{z}_2 (-\vk) + 2|\vk | \hat{q}_2(-\vk ).
\cr}\eqno(3.23)$$

Next, we wish to select a representation of $\ca$, ignoring for the moment
the $\star$-relations. Since we want the operators $\hat{z}_j(\vk )$ to act
by multiplication, we are led to choose for the carrier space $V$, the space
spanned by holomorphic distributions $\Psi(z_j(\vk ))$ of the type:
$$\eqalign{\Psi(z_j(\vk )) = h(z_j(\vk )) + \sum_{n=1}^N &\>
  \int \>{d^3k_1\over{|\vk_1|}} ...\lint{d^3k_n\over{|\vk_n|}}
  f_{j_1...j_n}(k_1, ... k_n)\cr
  &\times\>\>{\delta\over{\delta z_{j_1}(\vk_1)}}...
  {\delta\over{\delta z_{j_n}(\vk_n)}}\>\delta(z_1(\vk ))\delta(z_2(\vk ))\>,
  \cr}\eqno(3.24)$$
where $h(z_j(\vk ))$ is a holomorphic functional of $z_j(\vk )$ and where
the repeated indices $j_i$ are summed over $j_i=1,2$. Thus, at this
stage, without pre-judging the issue we allow both holomorphic functions and
derivatives of $\delta$-distributions for $j=1$ as well as $j=2$ modes. On
this $V$, the basic operators $\hat{q}_j(\vk )$ and $\hat{z}_j(\vk )$ are
represented by:
$$\eqalign{\hat{q}_i(-\vk )\cdot \Psi (z_j(\vk )) &= \hbar\>
{\delta\over{\delta z_i(\vk)}}\>\Psi (z_J(\vk )),\cr
\hbox{and}\quad \hat{z}_i(\vk ) \cdot \Psi (z_j(\vk )) &= z_i(\vk )
  \Psi (z_j(\vk ))\>,\cr}\eqno(3.25)$$
so that the CCR (3.22) are satisfied. The second of these equations ensures
that we are working in the self dual representation and the first then provides
the simplest way to achieve (3.22). Our job now is to select the inner product
using the reality conditions. The similarity of the two modes to the two
treatments of the harmonic oscillator enables us to follow the procedures
of sections 3.1 and 3.2 step by step. The logic of these calculations is
straightforward and care is needed only to keep track of which modes are
associated with momentum $\vk $ and which are associated with $-\vk $.
Therefore, we shall simply report the results.

The inner product is again expressible as:
$$\IP{\Psi (z_j)}{\Phi(z_j)} = \lint\left[\Pi_j\> {\rm d\!I}z_j(\vk )
\wedge {\rm d\!I}\ovr{z}_j(\vk )\right]\> \mu(z_j(\vk ),\ovr{z}_j(\vk))\>
\ovr{\Psi (z_j)} \> \Phi(z_j)\>,\eqno(3.26)$$
where ${\rm d\!I}$ is the infinite dimensional exterior derivative
on the space spanned by $(z_j(\vk ), \ovr{z}_j(\vk ))$. The reality
conditions again lead to a (functional) differential equation. It has a
unique solution: Upto an overall constant multiplicative factor, the measure
is given by:
 $$\mu(z_j(\vk ),\ovr{z}_j(\vk))= \exp -\bigg[\sum_j{\textstyle{(-1)^j\over 4
 \hbar}}\lint {d^3\vk\over |\vk |}\>(z_j(\vk)+\ovr{z}_j(-\vk ))(z_j(-\vk )
 +\ovr{z}_j(\vk ))\bigg].\eqno(3.27)$$
A simple calculation shows that, as in section 3.1, the measure ``damps
correctly'' for $j=2$ modes so that the normalizable states are just
holomorphic functionals $h(z_2(\vk ))$ of
$z_2(\vk )$. This space is left
invariant by the entire algebra $\cas$. (If we use distributional states as
well
for these modes, the norm fails to be positive definite.) For the $j=1$ modes,
on the other hand, as in section 3.2, the only holomorphic functional
$h(z_1(\vk ))$ that is normalizable is the zero functional. The normalizable
states are all distributional in $z_1(\vk )$. Thus, a general
normalized state is a superposition of states of the form:
\vfill\eject
$$\eqalignno{\Psi (z_j) =\lint {d^3\vk_1\over |\vk |_1}\> ...&\> \lint
  {d^3\vk_n\over |\vk |_n}\>f(\vk_1, ...\vk_n)\>\>
  {\delta\over{\delta z_1(\vec k_1)}}
  \> ...\>{\delta\over{\delta z_1(\vec k_n)}}\> \delta(z_1(\vk))\cr
   &\times \> P(z_2(\vk ))\,\exp\bigg[{\textstyle{1\over
   4\hbar}}\lint {d^3\vk \over{|\vk|}} (z_2(\vk )z_2(-\vk
))\bigg],&(3.28)\cr}$$
where $P(z_2(\vk ))$ is a polynomial in $z_2(\vk )$. The norm of this state
is given by:
$$\eqalignno{|\!|\Psi(z_j)|\!|^2& = \big( \lint {d^3\vk_1\over |\vk_1 |}
\>...\>
  \lint {d^3k_n\over |k_n|}\>|f(\vk_1, ...\vk_n)|^2 \big)\times\cr
  &\big( \lint {\rm d\!I}z_2(\vk )\wedge {\rm d\!I} \ovr{z}_2(\vk )\>
  \exp\bigg[\>-{1\over 2\hbar}\lint {d^3\vk\over|\vk|}
  |z_2(\vk )|^2\bigg]\>|P(z_2(\vk ))|^2\big) .&(3.29)\cr}$$

Finally, to make contact with the Fock representation, let us write down the
explicit expressions of the annihilation operators, the vacuum state and the
Hamiltonian. As one might expect from sections 3.1 and 3.2, there is an
asymmetry between the two modes. The annihilation operators are now given by
$\hat{z_1}(\vk )$ for $j=1$ and by $(\hat{z}_2(\vk ))^\star$ for the $j=2$
modes. Consequently, there is also an asymmetry in the expression of the vacuum
state. The vacuum is given by $\Psi_0(z_j) = \delta(z_1(\vk ))
\times\>\exp\>[{\textstyle{1\over 2\hbar}}\lint d^3\vk |\vk |^{-1} z_2(\vk )
z_2(-\vk )]\>$. It is interesting to translate this expression back in terms of
the self dual connection ${}^+\!A_a^T(\vx )$ using (2.11). One finds:
$\Psi({}^+\!A) = \delta(z_1(\vk ))\times {\exp}\> Y({}^+\!A)$, where
$Y({}^+\!A)$ is just the $U(1)$-Chern-Simons action of the self dual
connection ${}^+\!A$. Thus, on the $j=2$ sector, not only is the exponential of
the Chern-Simons action a normalizable state, but it is in fact the
vacuum.
 \footnote{4}{In retrospect this is not totally surprising. The
  Hamiltonian density of the Maxwell field can indeed be written in the form
  of a product, ${}^+B^a {}^-B_a$. On the $j=2$ sector, ${}^-\hat{B}_a$ acts
  essentially as an annihilation operator. The state it kills is just
  the exponential of the Chern-Simons action.}
This comes about {\it only} because we have used the self dual representation.
On the $j=1$ sector, on the other hand, the exponential of the Chern-Simons
functional --being an ordinary holomorphic functional-- fails to be
normalizable; only the holomorphic {\it distributions} are normalizable in
this sector. Finally, the normal ordered Hamiltonian is:
$$\eqalign{\hat{H} &=- \hbar\lint d^3\vk\>\left(\> \hat{z}_1^\star(\vk )
  \hat{z}_1(\vk )\>+\hat{z}_2(\vk )\hat{z}_2^\star(\vk ) \right)\cr
  &=-\hbar\lint d^3k\> \Bigg[\>-(z_1(-\vk )+ 2\hbar |k|{\delta\over
  {\delta z_1(\vk )}})\>z_1(\vk ) + z_1(\vk )(-z_2(-\vk )+
  \hbar |k|{\delta\over{\delta z_2(\vk )}}) \>\Bigg]\cr}\eqno(3.30). $$
Using the argument given in the case of the harmonic oscillator, it is
straightforward to establish that, inspite of this apparent asymmetry,  this
representation is unitarily equivalent to the Bargmann --and hence also the
Fock-- representation of free photons. The self dual representation is indeed
viable within the framework of the general program.

\goodbreak
\section{4}{Discussion}
The analysis of the last two section raises several interesting issues.
\item{1.}Perhaps the most surprising aspect of this analysis is the appearance
of holomorphic distributions. These appear to be indispensible to obtain a
representation in which the self dual connection is diagonal. Our treatment of
these distributions, however, is rather naive. Presumably there is a systematic
mathematical theory that underlies these ideas. It is likely that such a theory
would have other applications to quantum theory as well as to other areas of
physics.
\item{2.}It is encouraging that the self dual representation does in fact
exist for the Maxwell field because the conceptual ingredients used in its
construction are available also in general relativity [6-8]. Indeed, we
hve followed, step by step, a quantization program which was
introduced in the context of quantum general relativity. Furthermore, our
basic canonical variables are the direct analogs of the ones used in that
program. In general relativity, the corresponding self dual representation has
been used to address a number of conceptual as well as technical questions.
The conceptual problems include the possibility of the gravitational CP
violation and the issue of time [7]. An example of the technical progress
is the availability of exact solutions to quantum constraints in Bianchi IX
models [13]. The key assumption in these analyses is the existence of a
representation which is diagonal in the self dual (gravitational) connection
in which the {\it Hermitian} operator (analogous to) $\hat{E}^T_a(\vx )$ can be
expressed by a functional derivative. Apriori it is not obvious that these
assumptions are viable. That they in fact are viable in the well
understood Maxwell theory is therefore reassuring.
\item{3.} Nowhere in the construction did we carry out a decomposition of
fields into their positive and negative frequency parts. The two helicity
modes, labelled by $j$, arose as a technical by-product in the process of
implementing the quantization program, rather than as irreducible
representations of the Poincar\'e group. Indeed, we did not have to appeal
anywhere to the Poincar\'e invariance. Rather, it is the reality conditions
that selected for us the inner-product and the vacuum. Of course, had we not
restricted ourselves to Minkowski space, we may not have been able to carry out
the quantization program to completion. In this sense, the existence of the
Poincar\'e  invariance has presumably been used indirectly somewhere in the
construction. The observation is rather that since the group does not feature
in an {\it explicit} way anywhere, one can be hopeful that the program may be
successful in other contexts as well. This hope is borne out in 2+1 dimensional
quantum general relativity.
\item{4.} Finally, this example has taught us, rather clearly, an important
lesson about the quantization program: the actual imposition of the reality
conditions can be {\it quite} an involved procedure. Even after the elementary
classical variables are fixed, one may still have to invent new representations
of the algebra of quantum operators to ensure the existence of a sufficiently
large physical Hilbert space. There is as yet no systematic procedure available
to construct them. In particular, the self dual representation that we were led
to for the $j=1$ modes seems to fall outside the geometric quantization program
since the distributions involved do not appear to arise as polarized
cross-sections of a line bundle on the phase space. Is there perhaps a more
general framework that we can rely on for guidelines?
\bigskip

\goodbreak
{\bf Acknowledgements}

We wish to thank Vince Moncrief for raising some of the issues discussed here
and Ranjeet Tate and Lionel Mason for discussions on holomorphic distributions.
This work was supported by the NSF grants INT88-15209, PHY90-12099 and
PHY90-16733; by the research funds provided by Syracuse University; and, by a
SERC Visiting Fellowship (to AA).

\goodbreak
\section{}{References}
\item{[1]} R.~Penrose and W.~Rindler, {\it Spinors and space-time}, volumes 1
and 2 (Cambridge University Press, Cambridge, 1984).
\item{[2]} E.P.~Wigner, On unitary representations of the inhomogeneous Lorentz
group, Ann. Math. {\bf 40}, 149-204 (1939).
\item{[3]} A.~Ashtekar, A note on self duality and helicity, \JMP\ {\bf 27},
824-827 (1986).
\item{[4]} R.~Penrose, Nonlinear gravitons and curved twistor theory,
\GRG\ {\bf 7}, 31-52 (1976).
\item{[5]} M.~Ko, M.~Ludvigsen, E.T.~Newman and K.P.~Tod, The theory of
${\cal H}$-space, Phy. Rep. {\bf 71}, 51-139 (1981).
\item{[6]} A.~Ashtekar, New variables for classical and quantum gravity,
\PRL\ {\bf 57}, 2244-2247 (1986); New Hamiltonian formulation of general
relativity, \PR\ {\bf D36},
1587-1603 (1987).
\item{[7]} A.~Ashtekar, {\it Lectures on non-perturbative canonical
gravity} (Notes prepared in collaboration with R. Tate) (World Scientific,
Singapore, 1991).
\item{[8]} C.~Rovelli, Ashtekar formulation of general relativity and
 non-perturbative quantum gravity, \CQG\ (in press).
\item{[9]} A.~Ashtekar, C.~Rovelli and L.~Smolin, Gravitons and loops
(pre-print).
\item{[10]} A.~Ashtekar and A. Magnon, A geometrical approach to external
potential problems in quantum field theory, \GRG\ {\bf 12}, 205-223 (1980).
\item{[11]} V.~Bargmann, Remarks on a Hilbert space of analytic functions,
Proc. Natl. Acad. Sci.(U.S.A.) {\bf 48}, 199 (1962);
J.M.~Jauch, {\it Foundations of Quantum Mechanics}, (Addison Wesley, Reading,
1968), pages 215-219.
\item{[12]} H.~Kodama, Holomorphic wave function of the universe, \PR\
{\bf D42} 2548-2565 (1990).
\item{[13]} H.~Kodama, Specialization of Ashtekar's formalism to Bianchi
cosmology, Prog. Theo. Phys. {\bf 80}, 1024-1040 (1988); V.~Moncrief and
M.~Ryan, Amplitude-real-phase exact solutions for quantum mixmaster
universes, (preprint).

\bye